\documentclass{article}

\usepackage{spconf,amsmath,graphicx}
\usepackage{booktabs}
\usepackage{multirow}
\usepackage{bm}
\usepackage{latexsym}
\usepackage{newtxtext}
\usepackage[varg]{newtxmath}
\usepackage{url}
\usepackage{subfigure}
\usepackage{CJKutf8}

\newcommand{\Table}[1]{Table \ref{tb:#1}} 
\newcommand{\Fig}[1]{Fig. \ref{fig:#1}} 
\newcommand{\jp}[1]{ \begin{CJK}{UTF8}{ipxm}#1\end{CJK} }
\usepackage[
    backend=biber,
    style=ieee,
    maxbibnames=3,
    maxcitenames=3,
    doi=false,isbn=false,url=false,eprint=false
]{biblatex} 

\title{Cross-Dialect Text-To-Speech in Pitch-Accent Language \\ Incorporating Multi-Dialect Phoneme-Level BERT}

\name{Kazuki Yamauchi, Yuki Saito, Hiroshi Saruwatari}
\address{The University of Tokyo, Japan}

\begin{document}

\setlength{\abovedisplayskip}{3pt} 
\setlength{\belowdisplayskip}{3pt} 
\setlength\floatsep{5pt} 
\setlength\intextsep{5pt} 
\setlength\textfloatsep{5pt} 
\setlength{\dbltextfloatsep}{5pt} 
\setlength{\dblfloatsep}{3pt}
\maketitle

\begin{abstract}
We explore {\it cross-dialect text-to-speech (CD-TTS),} a task to synthesize learned speakers' voices in non-native dialects, especially in pitch-accent languages.
CD-TTS is important for developing voice agents that naturally communicate with people across regions. We present a novel TTS model comprising three sub-modules to perform competitively at this task. We first train a backbone TTS model to synthesize dialect speech from a text conditioned on phoneme-level accent latent variables (ALVs) extracted from speech by a reference encoder. Then, we train an ALV predictor to predict ALVs tailored to a target dialect from input text leveraging our novel multi-dialect phoneme-level BERT. We conduct multi-dialect TTS experiments and evaluate the effectiveness of our model by comparing it with a baseline derived from conventional dialect TTS methods. The results show that our model improves the dialectal naturalness of synthetic speech in CD-TTS.
\end{abstract}


\vspace{-3pt}
\section{Introduction}
\vspace{-3pt}

\begin{figure}[t]
  \centering
  \includegraphics[width=0.9\linewidth]{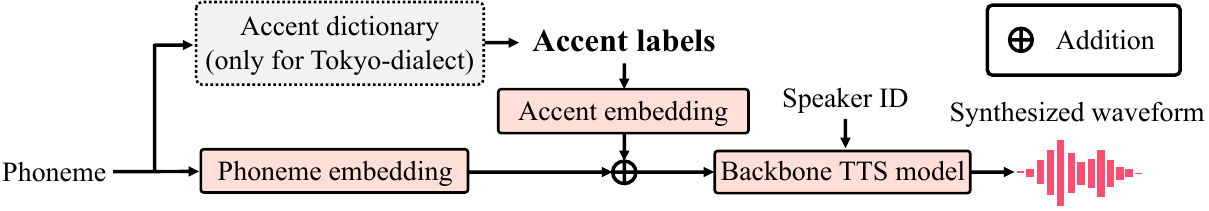}
  \vspace{-13pt}
  \caption{Flowchart of typical Japanese TTS model.}
  \label{fig:japanese_tts}
\end{figure}

Pitch-accent is a crucial prosodic attribute for natural speech communication in pitch-accent languages.
In Japanese, one of pitch-accent languages, each mora has its corresponding high or low (H/L) pitch-accent to distinguish homophones.
For instance, both\jp{雨}(rain) and\jp{飴}(candy) have the same pronunciation\jp{あめ}(a-me), but their pitch-accents (``HL'' and ``LH'') distinguish these words in Tokyo-dialect.
Therefore, typical Japanese text-to-speech (TTS)~\cite{sagisaka88} models take as input accent labels obtained using accent dictionaries (\Fig{japanese_tts}).

Dialects in pitch-accent languages each have a different pitch-accent rule.
For instance, in Osaka-dialect, one of Japanese dialects, ``LH'' pitch-accent is used to pronounce\jp{雨}(rain).
Therefore, it is essential for dialect TTS in pitch-accent languages to reproduce the pitch-accent of synthetic speech tailored to each dialect to avoid miscommunication.
However, building accent dictionaries to obtain the accent labels corresponding to texts for various dialects is very costly.
Indeed, accent dictionaries are only available for Tokyo-dialect in Japanese.
Therefore, current TTS systems find it challenging to adapt the pitch-accent of synthetic speech to different dialects, and this challenge is not well explored.

\begin{figure}[t]
  \centering
  \includegraphics[width=\linewidth]{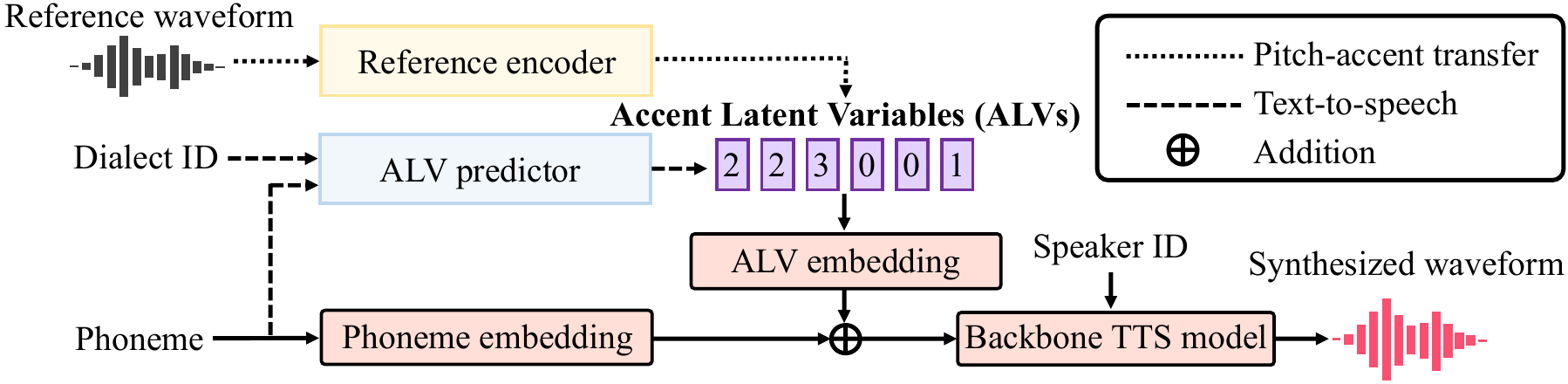}
  \vspace{-25pt}
  \caption{Overview of our proposed TTS model.}
  \label{fig:model_overview1}
\end{figure}

In this paper, we explore a new task called {\it cross-dialect (CD)-TTS}, which aims to synthesize learned speakers' voices in a non-native dialect, especially in pitch-accent languages.
CD-TTS is important for localizing TTS systems by adapting the pitch-accent of synthetic speech to regional dialects, leading to natural speech communication between computers and humans across regions.
Note that CD-TTS differs from existing {\it cross-lingual TTS}~\cite{xin21icassp}; specifically, CD-TTS focuses on several dialects within one specific language, which have similar but typically different pitch-accent systems and vocabularies.
We propose a novel TTS model for CD-TTS as illustrated in \Fig{model_overview1}, incorporating data-driven pitch-accent modeling using phoneme-level accent latent variables (ALVs).
Our model can automatically predict ALVs tailored to each dialect instead of relying on accent dictionaries.
Also, the ALV predictor incorporates a dialect-adapted version of phoneme-level BERT (PL-BERT)~\cite{li23plbert}, multi-dialect (MD)-PL-BERT, to improve the accuracy of ALV prediction.
The MD-PL-BERT is pre-trained on our constructed multi-dialect text corpus to capture both common and distinct textual features across dialects.
We conduct Japanese multi-dialect TTS experiments and compare our model with a baseline derived from conventional dialect TTS methods.
Audio samples are available on our demo page~\footnote{\url{https://kyamauchi1023.github.io/yamauchi24slt}}.
Our main contributions are as follows:
\begin{itemize}
    \item We explore a new task denoted as CD-TTS to synthesize learned speakers' voices in a non-native dialect.
    \item We propose a novel TTS model for CD-TTS that automatically predicts ALVs tailored to each dialect from text, leveraging our novel MD-PL-BERT.
    \item We present the result of evaluation experiments and demonstrate that leveraging our ALV predictor improves the dialectal naturalness of synthetic speech in CD-TTS.
\end{itemize}

\vspace{-3pt}
\section{Related Work}
\vspace{-3pt}

\vspace{-3pt}
\subsection{Prosody transfer}
\vspace{-3pt}
Prosody transfer~\cite{skerry-ryan18,robustfgpt,finegrainedpt,karlapati22_interspeech} is a technology to adapt prosody of synthetic speech to match that of reference speech while maintaining the speaker's voice timbre.
Typical prosody transfer methods extract speaker-independent latent representation of prosody from speech by a variational autoencoder (VAE)~\cite{kingma2014vae}-based reference encoder.
For example, Accent-VITS~\cite{accentvits}, a prosody transfer method for Chinese accented speech synthesis, extracts bottleneck (BN) features as prosody features from pre-trained automatic speech recognition (ASR) model and encodes them into latent representation by a VAE encoder.

\vspace{-3pt}
\subsection{Data-driven pitch-accent modeling}
\vspace{-3pt}
To address the challenge of Japanese dialect TTS, caused by the absence of accent dictionaries, Yufune et al.~\cite{yufune21} proposed a TTS method utilizing ALVs, instead of accent labels.
They first trained vector-quantized (VQ)-VAE~\cite{Oord2017VQVAE} to extract mora-level quantized latent representation from prosody features such as fundamental frequency (F0) of speech.
Since the representation can be regarded as pseudo accent label, they defined it as ALV.
They showed that VQ-VAE was more efficient than VAE used in typical prosody transfer methods for accurately reproducing the natural pitch-accent of synthetic speech in Japanese.
Then, they trained a TTS model conditioned on ALVs.
Also, they trained an ALV predictor that takes an input text and predicts ALVs corresponding to each mora.

\vspace{-3pt}
\subsection{Self-supervised pre-training on text data for TTS}
\vspace{-3pt}
It has been demonstrated that leveraging self-supervised pre-training on text data, such as PnG BERT~\cite{pngbert} and PL-BERT~\cite{li23plbert}, effectively improves the prosodic naturalness of synthetic speech by TTS.
PnG BERT is pre-trained on text data in a self-supervised manner, taking phonemes and graphemes of text as input.
PL-BERT, on the other hand, does not take graphemes as input; instead, it is pre-trained to predict graphemes from phonemes, aiming to enhance the robustness of prosody prediction for unknown graphemes not present in the training data.
In the context of Japanese Tokyo-dialect TTS, Japanese PnG BERT~\cite{japanesepngbert} improves the naturalness of pitch-accent of synthetic speech by pre-training to predict accent labels obtained using accent dictionaries.

\vspace{-3pt}
\subsection{Problems of conventional methods for dialect TTS}
\vspace{-3pt}

Yufune et al.'s study~\cite{yufune21} focused on single-speaker intra-dialect TTS (ID-TTS), i.e., synthesizing speech in the same dialect as the target speaker's native dialect.
Indeed, their model does not contain the functions to predict pitch-accent tailored to different dialects or adapt pitch-accent of synthetic speech to match that of an arbitrary speaker's reference speech.
Also, while they demonstrated that the naturalness of speech synthesized using ALVs extracted from ground-truth speech has improved, the naturalness of speech synthesized using predicted ALVs was lower than that of speech synthesized without ALVs, due to the low accuracy of ALV prediction.
Note that it has been demonstrated that inaccurate accent labels generally degrade the naturalness of synthetic speech~\cite{fujii22}.
One possible reason for the low ALV prediction accuracy of their model is the limited size of existing Japanese dialect speech corpora (e.g., CPJD~\cite{takamichi18cpjd}), which restricts the available data for training.
However, constructing speech corpora with a sufficient amount of data for each dialect is very costly.
Therefore, a method to improve the accuracy of ALV prediction without relying on additional dialect speech corpora is demanded.

Self-supervised pre-training on text data can be expected to improve the naturalness of pitch-accent for dialect TTS.
However, current text pre-training methods for TTS~\cite{japanesepngbert} typically utilize texts written in the standard language as training data, lacking mechanisms to learn features that vary across dialects.
Moreover, the availability of text corpora annotated with the dialect ID remains limited in size.
Therefore, a self-supervised pre-training method that is effective for dialect pitch-accent prediction is demanded for multi-dialect TTS.

\begin{figure*}[t]
  \centering
  \includegraphics[width=\linewidth]{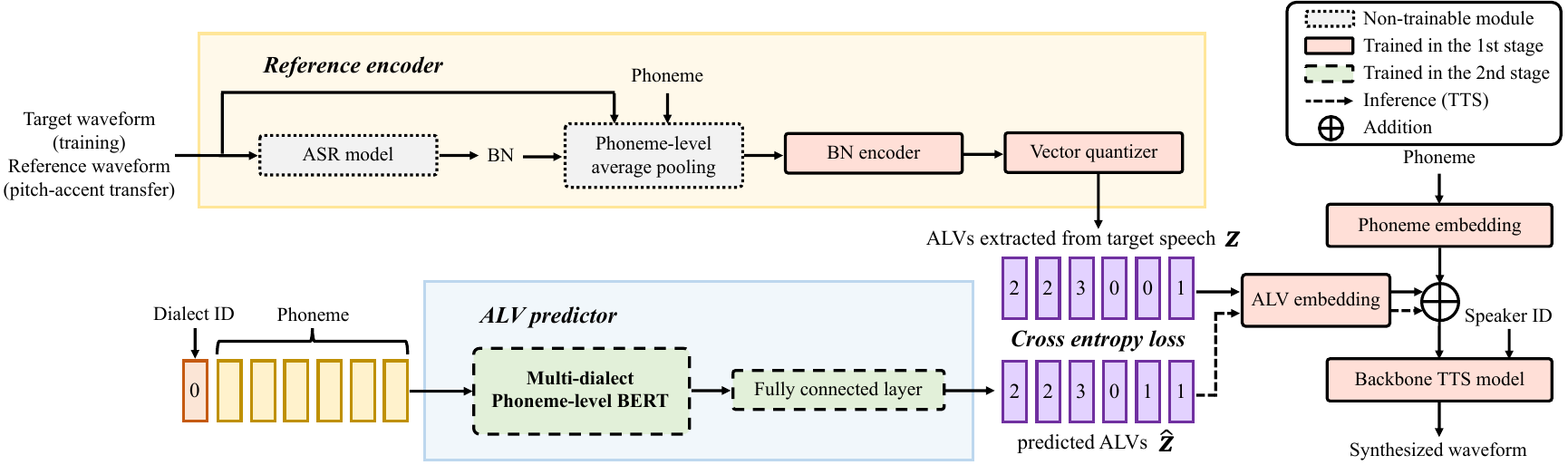}
  \caption{The architecture of our proposed model, consisting of a reference encoder and an ALV predictor. In the first training stage, the reference encoder and backbone TTS model are trained. In the second training stage, the ALV predictor is trained.}
  \label{fig:model_overview2}
\end{figure*}

\vspace{-3pt}
\section{Method}
\vspace{-3pt}
\Fig{model_overview2} illustrates the architecture of the proposed dialect TTS model, comprising: 1) a backbone TTS model, 2) a reference encoder, and 3) an ALV predictor.
The backbone TTS model synthesizes dialect speech conditioned on ALVs obtained by either of the other two modules.
The reference encoder extracts ALVs, phoneme-level quantized latent representation of prosody.
The ALV predictor predicts ALVs corresponding to each phoneme conditioned on a dialect ID.
The ALV predictor incorporates our novel MD-PL-BERT, pre-trained on our constructed multi-dialect text corpus, to capture both common and distinct textual features across dialects and to predict pitch-accent for phrases unique to each dialect.
Our model can synthesize speech from input text and a dialect ID by automatically predicting ALVs tailored to the target dialect (i.e., TTS).
Additionally, by inputting an arbitrary speaker's reference speech with the desired pitch-accent, the pitch-accent of the synthetic speech can be adapted to match that of the reference speech (i.e., pitch-accent transfer).

\vspace{-3pt}
\subsection{Reference encoder}
\vspace{-3pt}
The reference encoder is a module for extracting ALVs from prosody features of reference speech, enabling data-driven pitch-accent modeling without reliance on accent dictionaries.
We employ a VQ-VAE-based reference encoder, following Yufune et al.'s study~\cite{yufune21}.
Note that while Yufune et al. defined ALV at the mora-level~\cite{yufune21}, we define it at the phoneme-level.
To obtain prosody features related to pitch-accent information, the reference encoder incorporates a pre-trained ASR model into the ALV extraction framework, similar to the approach used in Accent-VITS~\cite{accentvits}.
Because pitch-accent is necessary for distinguishing words in pitch-accent languages, features obtained from a pre-trained ASR model are expected to contain sufficient prosody information.

Specifically, we first feed reference speech into the ASR model to extract BN features as the output of the ASR model's encoder's final layer.
BN features are aggregated into phoneme-level features using average pooling, guided by phoneme alignment information, to obtain phoneme-level ALVs.
Subsequently, they are fed into a one-dimensional convolutional neural network (1D-CNN)-based BN encoder.
Finally, the encoder outputs are quantized by a VQ module to obtain the quantized indices, i.e., the ALVs.

\vspace{-3pt}
\subsection{ALV predictor incorporating MD-PL-BERT}
\vspace{-3pt}

The ALV predictor is a module to predict ALVs tailored to a target dialect from input text.
We focus on PL-BERT~\cite{li23plbert}, a self-supervised learning model pre-trained on text data, to improve the accuracy of ALV prediction.
However, the original PL-BERT lacks mechanisms for learning linguistic features that vary across different dialects, making it challenging to predict ALVs specific to each dialect.
To address this, we propose MD-PL-BERT, a dialect-adapted version of PL-BERT, and incorporate it into the ALV predictor.
The pre-training strategy is similar to PL-BERT, but with two key differences.

First, we introduce conditioning PL-BERT on dialect ID, an identifier that indicates which dialect the input text is written in.
Specifically, we add a dialect ID to the beginning of the input phoneme sequence to enable PL-BERT to learn linguistic features tailored to the specified dialect.

Second, we construct a large-scale multi-dialect text corpus and pre-train MD-PL-BERT on them.
While pre-training MD-PL-BERT requires large-scale multi-dialect text corpora, the available text corpora annotated with dialect ID are limited in size.
Recent research has demonstrated the effectiveness of using large language models (LLMs) for dialect translation and has proposed a method for the automated construction of dialect text corpora~\cite{abdelaziz-etal-2024-llm}.
Inspired by this approach, we construct a multi-dialect text corpus by leveraging the data augmentation through translating texts written in the standard language (i.e., Tokyo-dialect in Japanese) into a target dialect using an LLM.
Specifically, we prompt a pre-trained LLM to translate a given Tokyo-dialect sentence into the target dialect using the following prompt: {\it ``Rewrite the following sentences as if they were in [target dialect]: [sentence written in Tokyo-dialect]''}.
The ALV predictor comprises MD-PL-BERT, pre-trained on this corpus, followed by a fully connected layer that predicts the ALVs from the output of the final layer of MD-PL-BERT.

\vspace{-3pt}
\subsection{Training and inference}
\vspace{-3pt}
Our model is trained in two stages.
In the first stage, the reference encoder and the backbone TTS model are jointly trained while the parameters of the pre-trained ASR model remain frozen.
The loss function is the sum of the losses from the backbone TTS model and the VQ loss~\cite{Oord2017VQVAE}.
During training, the target ground-truth speech is used as the reference speech.
In the second stage, the ALV predictor is initialized with the pre-trained MD-PL-BERT and fine-tuned together with a fully connected layer.
The loss function $\mathcal{L}$ used to train the ALV predictor is the cross-entropy loss (CELoss) between the ALVs extracted from the target speech by the reference encoder, $\bm{z}$, and the predicted ALVs, $\hat{\bm{z}}$, denoted as:
\begin{equation}
    \mathcal{L} = \mathrm{CELoss}(\bm{z}, \hat{\bm{z}})
\end{equation}

During inference, our TTS model enables pitch-accent transfer by synthesizing speech using ALVs extracted from an arbitrary speaker's reference speech.
This allows for control of the pitch-accent of synthetic speech by inputting reference speech with the desired pitch-accent.
Pitch-accent transfer can be seen as a variant of prosody transfer.
The key difference is that prosody transfer primarily focuses on emotion or speaking style, whereas pitch-accent transfer targets pitch-accent, which is discrete and more akin to linguistic information.

\vspace{-3pt}
\section{Experiments}
\vspace{-3pt}
We evaluate our method in both ID-TTS and CD-TTS.
The experiments focus on synthesizing speech in Osaka-dialect, one of Japanese dialects, by a native Osaka-dialect speaker (i.e., ID-TTS) and a Tokyo-dialect speaker (i.e., CD-TTS).
\vspace{-3pt}
\subsection{Experimental conditions}
\vspace{-3pt}

{\bf Training dataset:}
We used JSUT~\cite{jsut} and JMD~\footnote{\url{https://sites.google.com/site/shinnosuketakamichi/research-topics/jmd_corpus?authuser=0}}~\cite{jmd}.
JSUT consists of approximately 7,700 utterances by a single Tokyo-dialect speaker (female), while JMD includes 1,300 utterances by native dialect speakers for each dialect.
We mixed JSUT and the JMD-Osaka subset including voices by a single native Osaka-dialect speaker (female) and divided this mixed dataset into training (8,484 utterances), validation (256 utterances), and test (256 utterances) subsets.

{\bf Evaluation dataset:}
To evaluate the effectiveness of pitch-accent transfer using reference speech by an unseen speaker not present in the training dataset, we used speech in CPJD~\cite{takamichi18cpjd} as reference speech.
CPJD is a multi-dialect speech corpus collected through crowdsourcing, containing 250 utterances for each dialect.
We used the CPJD-Osaka subset including voices by a single native Osaka-dialect speaker (male) as reference speech for pitch-accent transfer.

{\bf Training setup:}
BN features were extracted by the encoder of the pre-trained Whisper large-v2 model\footnote{\url{https://huggingface.co/openai/whisper-large-v2}}~\cite{radford2022whisper}.
The phoneme alignment information to aggregate BN features into phoneme-level features was obtained using Julius~\cite{lee01julius}.
The BN encoder first projects BN features aggregated at the phoneme level into 256 dimensions and feeds them into a stack of two 1D-CNN layers with a kernel size of 3, stride of 1, and filter size of 256. 
This process outputs phoneme-level 256-dimensional continuous vectors.
Subsequently, the vectors are quantized into four classes, following the previous Japanese dialect TTS study~\cite{yufune21}.
Finally, the quantized vectors (i.e., ALV embeddings) are added to 256-dimensional phoneme embeddings.
Note that the indices of the quantized vectors are the ALVs.
The weight of the commitment loss in VQ loss~\cite{Oord2017VQVAE} was set to 4.0.
Also, we used FastSpeech~2~\cite{Ren2021FastSpeech2} as the backbone TTS model following the publicly available implementation (FastSpeech2-JSUT\footnote{\url{https://github.com/Wataru-Nakata/FastSpeech2-JSUT}}) for the network architecture and training settings.
That is, for the first stage of training, the model was trained with a batch size of 32, learning rate of 0.0625, and 100k iterations in 5 hours.
The pre-trained HiFi-GAN UNIVERSAL$\_$V1 model\footnote{\url{https://github.com/jik876/hifi-gan}} \cite{Jungil2020HiFiGAN} was used as a vocoder.

{\bf Pre-training:} For pre-training MD-PL-BERT, we used Japanese Wikipedia corpus\footnote{\url{https://dumps.wikimedia.org/}}, containing approximately 1.0M documents, and ReazonSpeech small\footnote{\url{https://huggingface.co/datasets/reazon-research/reazonspeech}}, containing approximately 62K utterances designed for building a Japanese ASR model.
We used transcriptions in ReazonSpeech as text dataset written in Tokyo-dialect and translate them into Osaka-dialect.
MD-PL-BERT was initialized by PL-BERT pre-trained on Wikipedia corpus and then pre-trained on transcriptions in ReazonSpeech with the data augmentation described in Section~3.2.
We used Japanese Llama~2~\cite{2023llama2}\footnote{\url{https://llama.meta.com/}}, a.k.a., 
Swallow~13B\footnote{\url{https://huggingface.co/tokyotech-llm/Swallow-13b-instruct-hf}} as the LLM for dialect translation.
We followed the network architecture and pre-training strategy of PL-BERT described in the official implementation (PL-BERT\footnote{\url{https://github.com/yl4579/PL-BERT}}).
To tokenize Japanese text into subwords, we used a publicly available tokenizer\footnote{\url{https://huggingface.co/tohoku-nlp/bert-base-japanese-whole-word-masking}}.
For grapheme-to-phoneme (G2P) conversion, we used OpenJTalk\footnote{\url{https://open-jtalk.sp.nitech.ac.jp}}.
PL-BERT was pre-trained on Wikipedia corpus with a batch size of 8, learning rate of $4.0\times10^{-6}$, and 10M iterations in 10 days.
MD-PL-BERT was pre-trained with a batch size of 16, learning rate of $5.0\times10^{-5}$, and 100k iterations in 10 hours.
For the second stage of training the proposed model, it was trained with a batch size of 32, learning rate of 0.001, and 10k iterations in 5 hours.

{\bf Model parameters and computational resources:} The backbone TTS model, the reference encoder, and the ALV predictor contained 35M, 790K, and 6M trainable parameters, respectively.
All the models were trained on a single Nvidia A100 GPU using the Adam optimizer~\cite{kingma14adam} with the linear scheduler of learning rate with warm up steps of 4000.

{\bf Task definition and compared models:}
We evaluated our proposed model through two tasks: 1) ID-TTS and 2) CD-TTS.
The former and latter aim to synthesize speech 1) in the same dialect as the target speaker's native dialect and 2) in a different dialect from the target speaker's native dialect, respectively.
The target speakers for ID-TTS and CD-TTS were defined as the JMD-Osaka speaker and the JSUT speaker, respectively.
Input texts for TTS are sampled from transcriptions in CPJD-Osaka.
We mainly evaluated the following models:
\begin{itemize} \leftskip -1mm \itemsep -0mm
\item {\bf FS2 (baseline)}: The original FastSpeech~2
\item {\bf FS2-AP (proposed)}: The proposed model using ALVs predicted by the ALV predictor form input text
\item {\bf FS2-REF (proposed)}: The proposed model using ALVs extracted from reference speech
\end{itemize}
\vspace{-2mm}

\vspace{-3pt}
\subsection{Evaluations}
\vspace{-3pt}
We conducted subjective and objective evaluations to compare the proposed model with an existing baseline.

{\bf Mean opinion score (MOS) tests:}
We conducted MOS tests via crowdsourcing to assess the naturalness of speech and the dialectal naturalness (i.e., {\it dialectality}) of pitch-accent for each method.
Participants evaluated randomly selected synthetic speech samples by each method or natural speech samples in CPJD from two viewpoints: 1) naturalness (N-MOS) and 2) dialectality (D-MOS).
The former and latter mean whether 1) it sounds naturally human-like and 2) its pitch-accent sounds natural as Osaka-dialect, not Tokyo-dialect, on a 5-point scale from 1 (very unnatural) to 5 (very natural), respectively.
For both ID-TTS and CD-TTS, 35 native Japanese speakers evaluated 24 randomly presented speech samples.

{\bf Pairwise comparisons:}
We also conducted several preference AB tests on the naturalness and dialectality of synthetic speech to determine the appropriate baseline and for ablation studies of the proposed model.
Twenty listeners participated in the tests via crowdsourcing, and each listener evaluated ten pairs of synthetic speech samples.
In the following subsection, {\bf bold} values in the tables showing the results of the AB test indicate that significant differences are determined by a Student’s $t$-test at a $5\%$ significance level.

{\bf Speaker similarity:}
To verify that ALVs are speaker-independent, we measured the speaker similarity of synthetic speech to the target speaker's natural speech, using cosine similarity between x-vectors~\cite{snyder2018xvector} (SIM).
Specifically, we computed the mean of SIM between the averaged x-vector among all speech samples of the target speaker in the test set and the x-vector of each synthetic speech.
We obtained x-vectors using a pre-trained model\footnote{\url{https://github.com/sarulab-speech/xvector_jtubespeech}}.

\begin{table}[tb]
\centering
\caption{Results of comparing the performance of FS2-AP-Scratch and FS2 in ID-TTS.}
\vspace{2mm}
\label{tb:result0a}
\scalebox{0.92}{
  \begin{tabular}{c|cc} \hline
    A vs. B & Naturalness & Dialectality \\ \hline \hline
    FS2-AP-Scratch vs. FS2 & 0.250 vs. {\bf 0.750} & 0.227 vs. {\bf 0.773} \\ \hline
  \end{tabular}
}
\end{table}

\vspace{-3pt}
\subsection{Results and discussion}
\vspace{-3pt}

{\bf What is the appropriate baseline in this study?} As well as FS2, one can regard FS2-AP without initialing the model parameter on the basis of the MD-PL-BERT pre-training (i.e., FS2-AP-Scratch) as the candidate baseline.
The reason is that the model structure and two-stage training of FS2-AP-Scratch are similar to those used in Yufune et al.'s study~\cite{yufune21}.
Therefore, we first compared these two methods in a preference AB test in ID-TTS. As shown in Table~\ref{tb:result0a}, FS2 significantly outperformed FS2-AP-Scratch, indicating that the prediction performance of ALV predictor without pre-training on text datasets is poor and inaccurate ALV prediction makes the naturalness of synthetic speech even worse.
This result is consistent with the result of Yufune et al.'s study~\cite{yufune21} mentioned in Section~2.4.
From this result, we decided to use FS2 as the baseline to be compared with the proposed model.

{
\tabcolsep = 1mm
\begin{table}[tb]
\centering
\small
\caption{Results of MOS test with $95\%$ confidence interval and computed SIM. REF represents the reference speech. \textbf{Bold} values are significantly higher than those of FS2 according to the results of a student’s $t$-test at a $5\%$ significance level.}
\label{tb:result1}
\subtable[ID-TTS: Synthesis of Osaka-dialect speech by Osaka-dialect speaker]{
\label{tb:result1a}
  \begin{tabular}{cc|cc|c} \hline
    Method & Target speaker & N-MOS ($\uparrow$) & D-MOS ($\uparrow$) & SIM ($\uparrow$) \\ \hline \hline
    FS2 & JMD (Osaka) & 3.30 $\pm$ 0.12 & 3.22 $\pm$ 0.13 & 0.990 \\
    FS2-AP & JMD (Osaka) & 3.31 $\pm$ 0.13 & 3.26 $\pm$ 0.13 & 0.991 \\ \hline
    FS2-REF & JMD (Osaka) & 3.23 $\pm$ 0.12 & 3.30 $\pm$ 0.12 & 0.992 \\ \hline
    REF & CPJD (Osaka) & \textbf{3.89 $\pm$ 0.14} & \textbf{4.38 $\pm$ 0.09} & - \\ \hline
  \end{tabular}
}
\subtable[CD-TTS: Synthesis of Osaka-dialect speech by Tokyo-dialect speaker]{
\label{tb:result1b}
  \begin{tabular}{cc|cc|c} \hline
    Method & Target speaker & N-MOS ($\uparrow$) & D-MOS ($\uparrow$) & SIM ($\uparrow$) \\ \hline \hline
    FS2 & JSUT (Tokyo) & 3.57 $\pm$ 0.13 & 2.62 $\pm$ 0.13 & 0.990 \\
    FS2-AP & JSUT (Tokyo) & 3.52 $\pm$ 0.13 & {\bf 3.00 $\pm$ 0.15} & 0.990 \\ \hline
    FS2-REF & JSUT (Tokyo) & 3.58 $\pm$ 0.12 & \textbf{3.05 $\pm$ 0.14} & 0.990 \\ \hline
    REF & CPJD (Osaka) & \textbf{4.39 $\pm$ 0.10} & \textbf{4.32 $\pm$ 0.13} & - \\ \hline
  \end{tabular}
}
\end{table}
}

\begin{table}[tb]
\centering
\caption{Results of comparing the performance of FS2 and FS2-AP in CD-TTS by native Osaka-dialect speakers.}
\label{tb:result2}
\vspace{1mm}
\scalebox{1.0}{
  \begin{tabular}{c|cc} \hline
    A vs. B & Naturalness & Dialectality \\ \hline \hline
    FS2 vs. FS2-AP & 0.506 vs. 0.494 & 0.387 vs. {\bf 0.613} \\ \hline
  \end{tabular}
}
\end{table}

\begin{table}[t]
\centering
\caption{Results of comparing the performance of pitch-accent transfer by FS2-REF using BN feature and F0 as prosody features.}
\label{tb:result3}
\vspace{1mm}
\scalebox{1.0}{
  \begin{tabular}{c|cc} \hline
    A vs. B & Naturalness & Dialectality \\ \hline \hline
    F0 vs. BN & 0.400 vs. {\bf 0.600} & 0.424 vs. {\bf 0.576} \\ \hline
  \end{tabular}
}
\end{table}

\begin{table}[tb]
\centering
\caption{BLUE@4 and BERTScore between Osaka-dialect sentences and original (Saitama-dialect) sentences or sentences translated into Osaka-dialect. {\bf Bold} scores are better.}
\label{tb:eval_llm}
\vspace{2mm}
\scalebox{1.0}{
  \begin{tabular}{c|cc} \hline
    Text & BLEU@4 ($\uparrow$) & BERTScore ($\uparrow$) \\ \hline \hline
    Original & 0.370 & 0.873 \\
    Translated & {\bf 0.401} & {\bf 0.882} \\ \hline
  \end{tabular}
}
\end{table}

\begin{table}[!h]
\centering
\caption{Results of comparing the absence of data augmentation (DA) by LLM-based dialect translation in CD-TTS.}
\label{tb:result4}
\vspace{1mm}
\scalebox{1.0}{
  \begin{tabular}{c|cc} \hline
    A vs. B& Naturalness & Dialectality \\ \hline \hline
    w/o DA vs. w/ DA & 0.491 vs. 0.509 & 0.343 vs. {\bf 0.657} \\ \hline
  \end{tabular}
}
\end{table}

{\bf Can our models improve dialect TTS performance?}
\Table{result1} shows the results of MOS tests.
First, from the results of ID-TTS shown in \Table{result1a}, no significant difference in MOS was observed between FS2 and FS2-AP.
Meanwhile, pitch-accent transfer through reference speech input tended to improve D-MOS, although the improvement was not statistically significant.
Second, from the results of CD-TTS shown in \Table{result1b}, FS2-AP achieved significantly higher D-MOS than FS2.
This indicates that the ALV predictor learned typical accent representation of Osaka-dialect, and the proposed model was effective in improving the dialectality of synthetic speech in CD-TTS.
Furthermore, pitch-accent transfer through reference speech input (i.e., FS2-REF) significantly improved D-MOS compared to FS2.
Also, using ALVs extracted from reference speech by a different speaker from the target speaker did not degrade the speaker similarity to the target speaker.
This demonstrates that our model enables pitch-accent transfer through an unseen speaker's reference speech input.

{\bf Is the improvement significant for native Osaka-dialect speakers?}
We asked eight native Osaka-dialect speakers to evaluate the naturalness and dialectality of synthetic speech by FS2 and FS2-AP in a preference AB test.
As shown in Table~\ref{tb:result2}, FS2-AP significantly outperformed FS2 in dialectality, while maintaining the naturalness.
This result demonstrates the effectiveness of the proposed TTS model is perceivable for not only crowdsourced listeners but also native dialect speakers.

\vspace{-3pt}
\subsection{Ablation study}
\vspace{-3pt}

{\bf Are BN features effective for pitch-accent transfer?}
Instead of BN features, F0 can be used as a prosody feature for ALV extraction, similar to Yufune et al.'s method~\cite{yufune21}.
Therefore, we compared the two prosody features, BN and F0, in the preference AB tests.
To obtain speaker-independent prosody features, we normalized F0 in an utterance-wise manner.
In addition, we linearly interpolated unvoiced regions of F0 in the phoneme-level average pooling.
We used WORLD~\cite{morise16world} to extract F0 from speech.
Also, we set the target speaker to the JSUT speaker.
The evaluation results are shown in \Table{result3}.
From this table, BN significantly outperformed F0 in both evaluation cases, demonstrating the effectiveness of BN for ALV extraction.
One possible reason is that while F0 is an acoustic feature, BN features can be considered as linguistic features acquired through the ASR task.

\begin{figure}[t]
  \centering
  \includegraphics[width=0.9\linewidth]{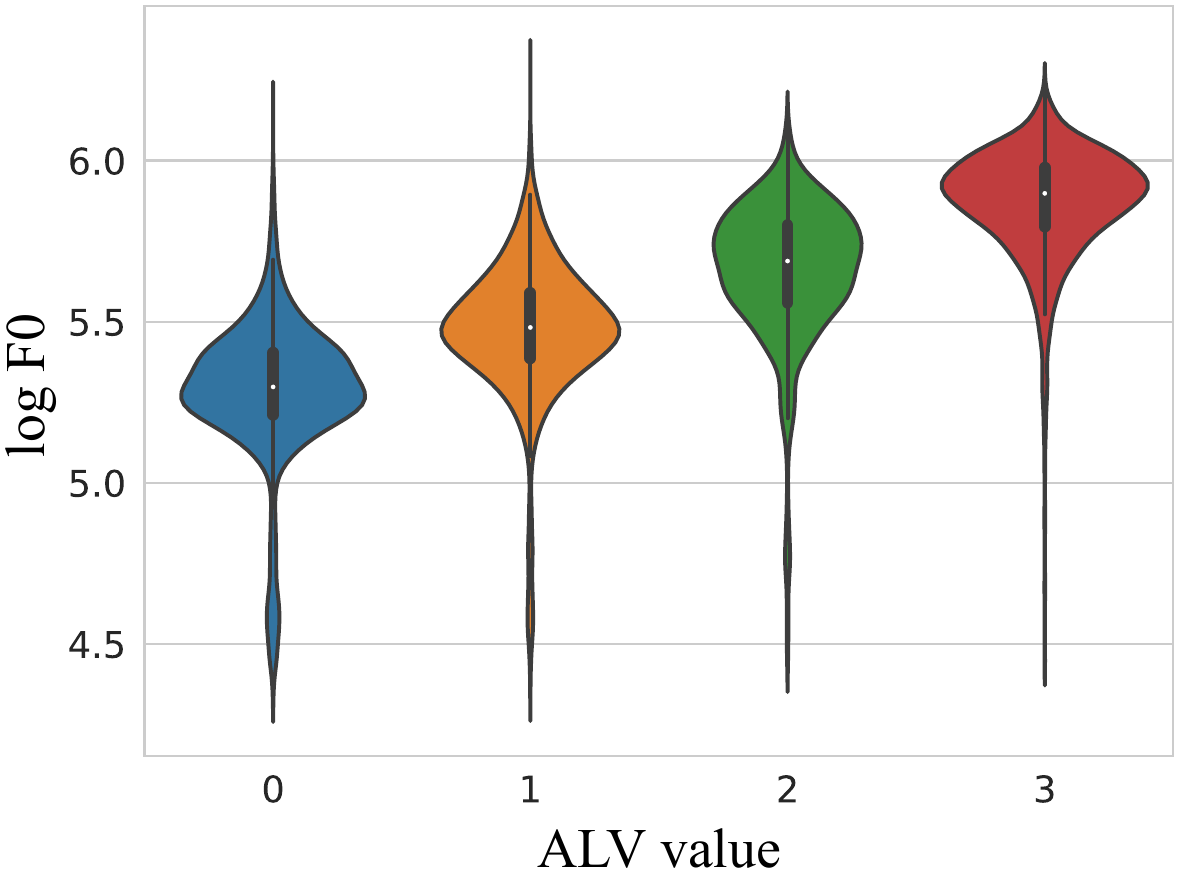}
  \vspace{-14pt}
  \caption{The violinplot of logarithmic fundamental frequency ($\log \mathrm{F0}$) aggrigated by ALV value ($0, 1, 2,$ or $3$).}
  \label{fig:pt-osaka_alv-logf0}
\end{figure}

{\bf How do ALVs influence the pitch-accent of synthetic speech?}
For TTS in pitch-accent languages, it is desirable that humans can easily correct errors in the pitch-accent of synthetic speech.
Therefore, we analyzed how ALVs influence the pitch-accent of synthetic speech to investigate the controllability and interpretability of ALVs.
Specifically, we extracted $\log$ F0 (logarithm of fundamental frequency) of synthetic speech and aggregated it at the phoneme-level.
The distribution was then plotted for each corresponding ALV.
We used FS2-REF with the target speaker being the JMD speaker and utilized the CPJD corpus as reference speech. The results are shown in \Fig{pt-osaka_alv-logf0}.
It can be observed that $\log$ F0 of synthetic speech varies according to ALV classes.
Specifically, $\log$ F0 for the intervals corresponding to ALV value $0 < 1 < 2 < 3$ tends to increase in order.
This suggests that ALVs can be interpreted as four categorical levels of pitch in synthetic speech and regarded as pseudo high-low pitch-accent labels.

{\bf Is the data augmentation by LLM-based dialect translation effective in improving the dialectality score?}
To verify the effectiveness of dialect translation by LLM as data augmentation, we initially conducted objective evaluations on translation accuracy.
We utilized transcriptions from CPJD, which contains semantically parallel transcriptions in multiple dialects.
Initially, we translated 250 transcriptions written in Saitama-dialect, the dialect closest to Tokyo-dialect within CPJD, into Osaka-dialect using an LLM.
Subsequently, we measured the similarity between the translated transcriptions and those originally written in Osaka-dialect in CPJD using BLEU~\cite{2002bleu} and BERTScore~\cite{2020bertscore}.
As shown in Table \ref{tb:eval_llm}, sentences translated by the LLM are more similar to Osaka-dialect than the original sentences.
This result indicates that the LLM has the ability for dialect translation.

We also conducted a subjective evaluation to assess the effectiveness of our MD-PL-BERT compared to the original PL-BERT.
Specifically, we compared two models in the preference AB tests: our FS2-AP incorporating MD-PL-BERT pre-trained with the data augmentation and the original PL-BERT pre-trained without the data augmentation.
As shown in Table~\ref{tb:result4}, our MD-PL-BERT, pre-trained on the multi-dialect text corpus constructed through the data augmentation, significantly improved the dialectality of synthetic speech.

\vspace{-3pt}
\section{Conclusions}
\vspace{-3pt}
We explored a new task called {\it cross-dialect text-to-speech (CD-TTS)}, which aims to synthesize learned speakers' voices in non-native dialects.
To address this, we proposed a novel TTS model comprising three sub-modules designed to perform effectively in this task.
We evaluated its performance not only on intra-dialect (ID)-TTS but also on CD-TTS through a series of subjective evaluations. 
The results show that our model improves the dialectality of synthetic dialect speech in CD-TTS without degrading the performance of ID-TTS.

In the future, we plan to investigate the effectiveness of our proposed model in dialect TTS using more dialects.
We also plan to incorporate machine learning techniques used to enhance the performance of cross-lingual TTS, such as domain adaptation~\cite{xin20interspeech} and mutual information minimization~\cite{xin21icassp}, into our model for CD-TTS.
Moreover, dialect TTS faces challenges not only with the lack of accent dictionaries but also with G2P converters.
Data-driven modeling of phoneme labels without reliance on G2P converters is also a future task.

\textbf{Acknowledgements:}
This work was supported by JST, ACT-X Grant Number JPMJAX23CB, Japan.

\printbibliography

@inproceedings{snyder2018xvector,
  author={D. Snyder and D. Garcia-Romero and G. Sell and D. Povey and S. Khudanpur},
  title={{X-Vectors}: Robust DNN Embeddings for Speaker Recognition},
  year=2018,
  booktitle={Proc. ICASSP},
  pages={5329--5333},
  month={Apr.},
  address={Calgary, Canada},
}

@inproceedings{2002bleu,
 author = {Kishore Papineni and Salim Roukos and Todd Ward and Wei-Jing Zhu},
 booktitle = {Proc. ACL},
 title = {{BLEU}: A method for automatic evaluation of machine translation},
 pages = {311--318},
 year = {2002},
 month = {Jun.},
 address = {Philadelphia, U.S.A.},
}

@article{2023llama2,
  title={{Llama 2}: Open Foundation and Fine-Tuned Chat Models},
  author={Hugo Touvron and Louis Martin and Kevin Stone and Peter Albert and Amjad Almahairi and Yasmine Babaei and Nikolay Bashlykov and Soumya Batra and Prajjwal Bhargava and Shruti Bhosale and Dan Bikel and Lukas Blecher and Cristian Canton Ferrer and Moya Chen and Guillem Cucurull and David Esiobu and Jude Fernandes and Jeremy Fu and Wenyin Fu and Brian Fuller and Cynthia Gao and Vedanuj Goswami and Naman Goyal and Anthony Hartshorn and Saghar Hosseini and Rui Hou and Hakan Inan and Marcin Kardas and Viktor Kerkez and Madian Khabsa and Isabel Kloumann and Artem Korenev and Punit Singh Koura and Marie-Anne Lachaux and Thibaut Lavril and Jenya Lee and Diana Liskovich and Yinghai Lu and Yuning Mao and Xavier Martinet and Todor Mihaylov and Pushkar Mishra, Igor Molybog and Yixin Nie and Andrew Poulton and Jeremy Reizenstein and Rashi Rungta and Kalyan Saladi and Alan Schelten and Ruan Silva and Eric Michael Smith and Ranjan Subramanian and Xiaoqing Ellen Tan and Binh Tang and Ross Taylor and Adina Williams and Jian Xiang Kuan and Puxin Xu and Zheng Yan and Iliyan Zarov and Yuchen Zhang and Angela Fan and Melanie Kambadur and Sharan Narang and Aurelien Rodriguez and Robert Stojnic and Sergey Edunov and Thomas Scialom},
  journal={arXiv preprint arXiv:2307.09288},
  year={2023}
}

@inproceedings{xin21icassp,
 address              = {Montreal, Canada},
 author               = {D. Xin and T. Komatsu and S. Takamichi and H. Saruwatari},
 booktitle            = {Proc. ICASSP},
 month                = {Jun.},
 pages                = {6608--6612},
 title                = {Disentangled Speaker and Language Representations Using Mutual Information Minimization and Domain Adaptation for Cross-Lingual {TTS}},
 year                 = {2021},
}

@inproceedings{xin20interspeech,
 address              = {Shanghai, China},
 author               = {D. Xin and Y. Saito and S. Takamichi and T. Koriyama and H. Saruwatari},
 booktitle            = {Proc. INTERSPEECH},
 month                = {Oct.},
 pages                = {2947--2951},
 title                = {Cross-Lingual Text-To-Speech Synthesis via Domain Adaptation and Perceptual Similarity Regression in Speaker Space},
 year                 = {2020},
}

@inproceedings{li23plbert,
  author    	      = {Y. A. Li and C. Han and X. Jiang and N. Mesgarani},
  title     		  = {Phoneme-Level {BERT} for Enhanced Prosody of Text-to-Speech with Grapheme Predictions},
  booktitle         = {Proc. ICASSP},
  year      		  = {2023},
  address           = {Rhodes, Greece},
  month             = {Jun.},
 }

@inproceedings{robustfgpt,
 author               = {Y. Lee and T. Kim},
 booktitle            = {Proc. ICASSP},
 pages                = {5911--5915},
 title                = {Robust and fine-grained prosody control of end-to-end speech synthesis},
 year                 = {2019},
 month                = {May},
 address              = {Brighton, U.K.},
}

@inproceedings{finegrainedpt,
  author			  = {V. Klimkov and S. Ronanki and J. Rohnke and T. Drugman},
  title				  = {Fine-grained robust prosody transfer for single-speaker neural text-to-speech},
  year 				  = {2019},
  booktitle			  = {Proc. INTERSPEECH},
  pages				  = {4440--4444},
  address               = {Graz, Austria},
  month                 = {Sep.},
}

@inproceedings{karlapati22_interspeech,
  author={Sri Karlapati and Penny Karanasou and Mateusz Łajszczak and Syed {Ammar Abbas} and Alexis Moinet and Peter Makarov and Ray Li and Arent {van Korlaar} and Simon Slangen and Thomas Drugman},
  title={{CopyCat2}: A Single Model for Multi-Speaker TTS and Many-to-Many Fine-Grained Prosody Transfer},
  year=2022,
  booktitle={Proc. INTERSPEECH},
  pages={3363--3367},
  month={Sep.},
  address={Incheon, South Korea},
}

@inproceedings{accentvits,
  author={L. Ma and Y. Zhang and X. Zhu and Y. Lei and Z. Ning and P. Zhu and L. Xie},
  title={{Accent-VITS}: accent transfer for end-to-end {TTS}},
  year=2023,
  booktitle={Proc. NCMMSC},
  month={Dec.},
  address={Suzhou, China},
}

@article{jmd,
 author    			  = {S. Takamichi and H. Saruwatari},
  title     		  = {{JMD}: Japanese multi-dialect corpus},
  year      		  = {2021},
  url       		  = {https://sites.google.com/site/shinnosuketakamichi/research-topics/jmd_corpus?authuser=0},
 }

@inproceedings{Oord2017VQVAE,
 author = {A. van den Oord and O. Vinyals and K. Kavukcuoglu},
 booktitle = {Proc. NIPS},
 pages = {6309--6318},
 title = {Neural Discrete Representation Learning},
 year = {2017},
 address={Long Beach, U.S.A.},
 month={Dec.}
}

@inproceedings{fujii22,
 address              = {Chiang Mai, Thailand},
 author               = {K. Fujii and Y. Saito and H. Saruwatari},
 booktitle            = {Proc. APSIPA ASC},
 month                = {Nov.},
 pages                = {1702--1707},
 title                = {Adaptive End-to-End Text-to-Speech Synthesis Based on Error Correction Feedback from Humans},
 year                 = {2022},
 }

@inproceedings{yufune21,
 address              = {Budapest, Hungary},
 author               = {K. Yufune and T. Koriyama and S. Takamichi and H. Saruwatari},
 booktitle            = {Proc. SSW},
 month                = {Aug.},
 pages                = {189--194},
 title                = {Accent modeling of low-resourced dialect in pitch accent language using variational autoencoder},
 year                 = {2021},
 }

@inproceedings{lee01julius,
 address              = {Aalborg, Denmark},
 author               = {A. Lee and T. Kawahara and K. Shikano},
 booktitle            = {Proc. EUROSPEECH},
 month                = {Sep.},
 pages                = {1691--1694},
 title                = {Julius --- An Open Source Real-Time Large Vocabulary Recognition Engine},
 year                 = {2001},
 }

@article{morise16world,
 author               = {M. Morise and F. Yokomori and K. Ozawa},
 journal              = {IEICE Transactions on Information and Systems},
 pages                = {1877--1884},
 title                = {{WORLD}: a vocoder-based high-quality speech synthesis system for real-time applications},
 number               = {7},
 volume               = {E99-D},
 year                 = {2016},
 month                = {Jul.}
 }

@inproceedings{sagisaka88,
 address              = {New York, U.S.A.},
 author               = {Y. Sagisaka},
 booktitle            = {Proc. ICASSP},
 issn                 = {1520-6149},
 month                = {Apr.},
 pages                = {679--682},
 title                = {Speech Synthesis by Rule Using an Optimal Selection of Non-uniform Synthesis Units},
 year                 = {1988},
 }

@article{jsut,
  title={{JSUT} and {JVS}: Free {J}apanese Voice Corpora for Accelerating Speech Synthesis Research},
  author={Shinnosuke Takamichi and Ryosuke Sonobe and Kentaro Mitsui and Yuki Saito and Tomoki Koriyama and Naoko Tanji and Hiroshi Saruwatari},
  journal={Acoustical Science and Technology},
  volume={41},
  number={5},
  pages={761-768},
  year={2020},
  month={Sep.},
}

@inproceedings{kingma2014vae,
  author    		  = {D. P. Kingma and M. Welling},
  booktitle         = {Proc. ICLR},
  title     		  = {Auto-Encoding Variational {Bayes}},
  year      		  = {2014},
  address           = {Banff, Canada},
  month             = {Apr.},
}

@inproceedings{takamichi18cpjd,
 address              = {Miyazaki, Japan},
 author               = {S. Takamichi and H. Saruwatari},
 booktitle            = {Proc. LREC},
 month                = {May},
 pages                = {434--437},
 title                = {{CPJD Corpus}: Crowdsourced Parallel Speech Corpus of {Japanese} Dialects},
 year                 = {2018},
 }

@inproceedings{kingma14adam,
  author    = {Diederik P. Kingma and
               Jimmy Ba},
  title     = {Adam: {A} Method for Stochastic Optimization},
  booktitle = {Proc. ICLR},
  year      = {2015},
  address = {San Diego, California, U.S.A.},
  month = {May}
}

@inproceedings{Ren2021FastSpeech2,
title={{FastSpeech} 2: Fast and High-Quality End-to-End Text to Speech},
author={Yi Ren and Chenxu Hu and Xu Tan and Tao Qin and Sheng Zhao and Zhou Zhao and Tie-Yan Liu},
booktitle={Proc. ICLR},
year={2021},
address={Vienna, Austria},
month={May}
}

@inproceedings{Jungil2020HiFiGAN,
 author = {Kong, Jungil and Kim, Jaehyeon and Bae, Jaekyoung},
 booktitle = {Proc. NeurIPS},
 pages = {17022--17033},
 title = {{HiFi-GAN}: Generative Adversarial Networks for Efficient and High Fidelity Speech Synthesis},
 year = {2020},
 address={Virtual Conference},
 month={Dec.}
}

@InProceedings{skerry-ryan18,
  title = 	 {Towards End-to-End Prosody Transfer for Expressive Speech Synthesis with {Tacotron}},
  author =       {Skerry-Ryan, RJ and Battenberg, Eric and Xiao, Ying and Wang, Yuxuan and Stanton, Daisy and Shor, Joel and Weiss, Ron and Clark, Rob and Saurous, Rif A.},
  booktitle = {Proc. ICML},
  pages = 	 {4693--4702},
  year = 	 {2018},
  month = 	 {July},
}

@inproceedings{2020bertscore,
  title={{BERTScore}: Evaluating Text Generation with {BERT}},
  author={Tianyi Zhang and Varsha Kishore and Felix Wu and Kilian Q. Weinberger and Yoav Artzi},
  booktitle={Proc. ICLR},
  year={2020},
  address={Virtual Conference},
  month={Apr.},
}

@inproceedings{radford2022whisper,
    author = {Radford, Alec and Kim, Jong Wook and Xu, Tao and Brockman, Greg and McLeavey, Christine and Sutskever, Ilya},
    title = {Robust Speech Recognition via Large-Scale Weak Supervision},
    booktitle = {Proc. ICML},
    year = {2023},
    pages = {28492--28518},
    month = {Jun.},
    address = {Hawaii, U.S.A.}, 
}

@article{japanesepngbert,
 author         = {Y. Yasuda and T. Toda},
 journal        = {IEEE Journal of Selected Topics in Signal Processing},
 number         = {6},
 pages          = {1319--1328},
 title          = {Investigation of {Japanese PnG BERT} language model in text-to-speech synthesis for pitch accent language},
 volume         = {16},
 year           = {2022},
 }

@inproceedings{pngbert,
  author={Y. Jia and H. Zen and J. Shen and Y. Zhang and Y. Wu},
  title={{PnG BERT: Augmented BERT on phonemes and graphemes for neural TTS}},
  year={2021},
  booktitle={Proc. INTERSPEECH},
  pages={151--155},
}

@inproceedings{abdelaziz-etal-2024-llm,
    title = "{LLM}-based {MT} Data Creation: Dialectal to {MSA} Translation Shared Task",
    author = "Abdelaziz, AhmedElmogtaba Abdelmoniem Ali  and
      Elneima, Ashraf Hatim  and
      Darwish, Kareem",
    booktitle = "Proc. OSACT Workshop",
    month = "May.",
    year = "2024",
    address = "Torino, Italia",
    url = "https://aclanthology.org/2024.osact-1.14",
    pages = "112--116",
}

\end{document}